\documentclass[12pt]{article}
\usepackage{latexsym}
\usepackage{graphicx}
\usepackage{mathptmx}
\usepackage[T1]{fontenc}
\usepackage[margin=1in]{geometry}
\usepackage[round,authoryear]{natbib}

\newcommand{\email}[1]{\texttt{#1}}

\usepackage{amsmath}
\usepackage{amsfonts}
\usepackage{amssymb}
\usepackage{amsbsy}
\usepackage{amsthm}
\usepackage{booktabs}
\usepackage[table]{xcolor}
\usepackage{booktabs}
\usepackage{colortbl}

\usepackage[colorlinks=true,urlcolor=blue,citecolor=black,linkcolor=black]{hyperref}

\newtheoremstyle{plainwsc}
{3pt}
{3pt}
{}
{}
{\bf}
{}
{.5em}
{}

\theoremstyle{plainwsc}

\newcommand{\js}[2]{\cellcolor{blue!#1}#2}
\newcommand{\improve}[2]{\cellcolor{blue!#1}#2}
\newcommand{\worsen}[2]{\cellcolor{red!#1}#2}
\newcommand{\meanval}[1]{\cellcolor{gray!15}\textbf{#1}}

\begin{document}

\title{Using Zero-Shot LLM-Generated Survey Data for Geographically Explicit Population Synthesis}

\author{
Taylor Anderson\textsuperscript{1},
Sara Von Hoene\textsuperscript{1},
Orhan Yagizer Cinar\textsuperscript{2},
Emma Von Hoene\textsuperscript{1},\\
Amira Roess\textsuperscript{3},
Andrew Crooks\textsuperscript{4},
and Hamdi Kavak\textsuperscript{5}\\[1em]
\small \textsuperscript{1}Dept.~of Geography and Geoinformation Science, George Mason University, Fairfax, VA, USA\\
\small \textsuperscript{2}Dept.~of Computer Science, George Mason University, Fairfax, VA, USA\\
\small \textsuperscript{3}College of Public Health, George Mason University, Fairfax, VA, USA\\
\small \textsuperscript{4}Dept.~of Geography, University at Buffalo, Buffalo, NY, USA\\
\small \textsuperscript{5}Dept.~of Computational and Data Sciences, George Mason University, Fairfax, VA, USA
}
\date{}

\maketitle

\section*{Abstract}
There is a growing interest in utilizing synthetic populations for a diverse range of applications. At the same time, we are witnessing a tremendous growth in artificial intelligence in all walks of life. This paper evaluates whether zero-shot large language model (LLM)-generated health survey data can serve as inputs to a conventional iterative proportional fitting (IPF) workflow for geographically explicit population synthesis. Using the 2023 Behavioral Risk Factor Surveillance System (BRFSS), we generate synthetic survey records for the U.S.\ states of Colorado and Mississippi with GPT-4.1 and Gemini-2.5-Pro. We use the generated data in an IPF-based synthesis pipeline and evaluate the resulting census tract-level synthetic populations against external benchmarks. Results show both LLMs capture several major state-level contrasts, indicating zero-shot generation produces geographically differentiated survey data. However, performance is strongly variable-dependent. Downstream effects in population synthesis are mixed, as IPF sometimes amplifies or reduces errors in the generated data. Spatial validation shows that LLM-based populations reproduce census tract-level patterns reasonably well, especially for variables that were more aligned with the ground truth data. Overall, the LLM-generated survey data shows promise as supplementary input, but not yet as a replacement for real survey data.

\section{Introduction}
\label{sec:intro}

Geographically explicit population synthesis is a technique for generating synthetic populations of individuals by integrating survey data with spatially aggregated marginal population constraints. These synthetic populations are widely used as inputs to agent-based and microsimulation models in public health, including models of infectious disease transmission \citep{grefenstette2013fred,kerr2021covasim}, chronic disease dynamics \citep{hennessy2015population,nascimento2025stress}, and the spread of health risk behaviors \citep{levy2010reaching,yin2024agent}. A common method for constructing them is iterative proportional fitting (IPF), which reweights or allocates survey-based records so they match marginal population totals for each geographic unit. 

The quality of a synthetic population depends not only on the fitting procedure itself, but also on the quality and availability of the input survey data. Bias, error, or sampling noise can propagate through the synthesis process and affect downstream spatial estimates and simulation outcomes. At the same time, many population synthesis applications are constrained by the limited availability of survey data that include the variables needed for a given simulation and/or are representative of the specific population being modeled.

One way to potentially overcome the gap in data is to utilize artificial intelligence. Recent work suggests that, with caution, large language models (LLMs) may be able to generate synthetic human data that resembles real survey respondents. For example, computational social science, studies have shown that LLMs can reproduce opinions and behaviors when prompted with background attributes or persona-like descriptions \citep{argyle2023out,marigliano2025aurora,westwood2025potential,bhandari2024urban,paglieri2026persona}. However, applications exploring LLM-generated survey data in the health domain remain more limited, although emerging research suggests that LLMs may be useful for generating synthetic health survey records \citep{villarreal2026generation}. Such work suggests that LLM-generated survey data could serve as inputs to geographically explicit population synthesis pipelines, addressing persistent data limitations.  

Other work is emerging that moves beyond individual record generation and attempts to scale LLM-produced individuals into larger synthetic populations. In some cases, this has involved specifying constraints in the prompt to align LLM-generated populations with known spatial distributions \citep{tang2025llmsynthor}. While Olmez et al. (\citeyear{olmez2025generating})  used LLMs to supplement or complete limited survey data before those data are used in population synthesis. Finally, some studies have also attempted to generate location itself as one of the individual attributes \citep{qin2026semapop}. This growing body of research  points towards a possible role for LLMs in synthetic population construction, but none of which fully address a different and more practical question for population synthesis: to what extent can LLM-generated survey data serve as suitable input data for established geographically explicit synthesis methods?

This distinction is important, as in geographically explicit population synthesis, the survey data are not the final product; they are the joint distributional template. As a result, the challenge is not only about the plausibility of LLM-generated responses in isolation, but about whether they preserve the multivariate relationships needed for downstream synthesis procedures such as IPF. To our knowledge, no work has directly examined the consequences of using LLM-generated survey data in this way. 

As such, in this paper, we take a first step toward evaluating the potential of zero-shot LLM-generated survey data for geographically explicit population synthesis using IPF. IPF makes for an interesting case study because it seeks to preserve the joint distributions present in the input records; thus, any incorrect relationships, bias, misclassification, or underrepresented subgroup patterns in the original data can be carried into the final synthetic population \citep{han2025quantitative}. To demonstrate our work, we focus on two substantively different U.S.\ states, Colorado (CO) and Mississippi (MS), and use fourteen categorical questions from the 2023 Behavioral Risk Factor Surveillance System or BRFSS health survey \citep{BRFSS}. These states provide a useful contrast because they differ across a range of demographic and health-related characteristics, including \textit{education}, \textit{income}, \textit{exercise}, \textit{body mass index}, and \textit{self-reported health}. We use the LLMs GPT-4.1 and Gemini-2.5-Pro to generate synthetic individual-level survey records and compare them with weighted BRFSS survey data to assess how well the models reproduce marginal distributions and broader state-level contrasts. We then use the generated records as inputs to an IPF-based population synthesis pipeline and evaluate the resulting census tract-level synthetic populations against external benchmarks.

Our contribution is twofold. First, whereas prior work has largely evaluated LLM-generated data at the national level, we assess the extent to which LLM-generated survey data can approximate state-specific health survey data and capture meaningful geographic differences. Second, we examine the downstream usefulness of these data within a geographically explicit population synthesis workflow by comparing the errors they propagate to those arising from a synthetic population generated from real survey data. Although our analysis considers all modeled variables, we use \textit{health insurance} and \textit{general health} as especially detailed case studies because they illustrate two different types of downstream behavior and can be validated against census tract-level external benchmarks. In what follows, we provide our overview of our methods (Section \ref{SEC:Methods} before presenting the results in Section \ref{SEC:Results}. Finally, we conclude our paper in Section \ref{SEC:Discussion} with a discussion and highlight areas of further work. 

\section{METHODS}\label{SEC:Methods}

Population synthesis fits survey data to spatially aggregated demographic data containing marginal population totals. Here, survey data refers to anonymized individual responses to a set of questions, including demographic information. In this study, for each state CO and MS, we generated populations using LLM-generated survey data from GPT-4.1, Gemini-2.5-Pro, and real BRFSS survey data, for a total of six populations. The following sections outline the methods for survey data generation using LLMs (Section \ref{SEC:Gen}), the population synthesis (Section \ref{SEC:PopSyn}), and our evaluation approach (Section \ref{SEC:Evaluation}).

\subsection{Survey data generation using LLMs}\label{SEC:Gen}
We generated synthetic individual-level survey records using two state-of-the-art LLMs at the time of this study: OpenAI's GPT-4.1 \citep{openai2026gpt} and Google's Gemini-2.5-Pro \citep{google2026gemini}. Our rationale for choosing two was to be able to compare and contrast them. In order to increase reproducibility and limit bias, we used their official application programming interfaces (APIs), rather than their web interface. For each combination of model and study area, we produced a data set that matches the size of the corresponding weighted BRFSS sample: 6,220 records for CO and 3,044 for MS, yielding four generated datasets in total.

Each LLM was prompted in a zero-shot setting, meaning no example survey records, target distributions, or numerical information were provided to the models. We instructed the LLMs to generate survey responses that are representative of the 2023 adult population of the two U.S. states. We further asked LLMs to use their knowledge of the Behavioral Risk Factor Surveillance System Survey or any other sources to make marginal and joint distributions as realistic and accurate as possible. We then specified 14 BRFSS variables and complete response coding, covering demographics (\textit{sex}, \textit{age}, \textit{education}, \textit{income}, \textit{race/ethnicity}), \textit{health insurance type}, \textit{self-rated general health}, chronic conditions (\textit{heart disease}, \textit{depression}, \textit{diabetes}), health behaviors (\textit{smoking}, \textit{exercise}, \textit{flu vaccination}), and \textit{BMI} category. 

As current LLMs cannot reliably generate thousands of structured records in a single API call, we implemented a batch generation pipeline that validated records across multiple sequential calls until the target sample size was reached. This batch-style approach has been used by others to generate synthetic survey data \citep{zhao2025large}. Each call requested 75 survey records and the batch size was selected based on experiments in which we tested sizes of 50, 75, 100, 150, and 200 rows per call. At larger batch sizes, we experienced output truncation and malformed JSON responses. This behavior is consistent with broader observations that directly prompting LLMs for structured data generation often produces repetitive outputs even at high temperatures \citep{long2024llms}. At 75 rows per call, these effects were minimal while maintaining practical generation. 

To guarantee programmatic adherence to the survey structure and eliminate formatting errors and hallucinations, we enforced a strict JSON schema via the models' structured output capabilities. We note that the structured output enforcement is syntactic, so LLMs can still occasionally generate integers outside the defined BRFSS response codes. Our subsequent row-level validation step therefore remained critical, checking each record against a predefined dictionary of valid response codes for every variable. Records with any out of range value were dropped rather than corrected, to avoid introducing bias through imputation. This approach aligns with the general data curation workflow for LLM-generated tabular data, in which filtering removes corrupted or out of distribution records after generation, a crucial step for maintaining high quality synthetic datasets \citep{long2024llms,seedat2024curated}.

Both models were configured with a temperature of 1.0 and top-p of 1.0 to maximize diversity in the generated responses. 
Frequency and presence penalties were set to 0.0 for models to avoid artificially suppressing repeated responses; repetition is expected in structured survey data and suppressing it would distort marginal distributions. Both models used a maximum output token limit of 32,768 per call. The pipeline included checkpoint based persistence, saving records and metadata to disk at regular intervals so that generation could resume from the most recent checkpoint. The code and prompts used for survey data 
generation are publicly available at \url{https://doi.org/10.17605/OSF.IO/H4STJ}.

\subsection{Population synthesis}\label{SEC:PopSyn}

We generated synthetic populations using IPF, a standard approach for constructing geographically explicit synthetic populations by aligning individual-level survey data will spatially aggregated demographic constraints \citep{LOVELACE20131,huang2001comparison,vonhoene2025}. In total, six synthetic populations are generated for adults aged 18 and older (three for each study area). The populations representing CO resulted in a total of 4,400,554 synthetic individuals, while the synthetic population of MS contained 2,156,911. For each state, one population is constructed using BRFSS survey data as a 'ground-truth' reference, while the remaining two are generated using LLMs. All populations are fitted to census tract-level aggregated demographic data (2019–2023 five-year estimates) from the U.S. Census Bureau’s American Community Survey (ACS) \citep{acs2023}, such that the differences across populations are driven solely by the underlying survey inputs. 

The synthetic populations are generated using \textit{age}, \textit{race/ethnicity}, \textit{gender}, \textit{income}, and \textit{education} as fitting/predictor variables. These characteristics are selected due to their well-established relationships with the health attributes investigated in this study \citep{stone2015place,kunnath2024relative}. Each population is constructed at the tract-level, with individuals assigned a “home” tract and the associated demographic attributes. 
Following standard IPF procedures using health survey data \citep{vonhoene2025}, survey records are iteratively reweighted to match marginal demographic totals for each tract while preserving joint distributions from the input data. After integerization and replication \citep{LOVELACE20131}, additional attributes from the input survey dataset \textit{body mass index (BMI)/obesity}, \textit{depression}, \textit{diabetes}, \textit{exercise behavior}, \textit{flu vaccination}, \textit{general health status}, \textit{heart disease}, \textit{health insurance coverage}, and \textit{smoking behavior} are 'carried over' or joined to each synthetic individual. This results in a geographically explicit, health-enriched synthetic population for each study area.

\subsection{Evaluation}\label{SEC:Evaluation}

We first assess the quality of the LLM-generated survey data by comparing their categorical variable distributions to the 2023 BRFSS ground truth survey data \citep{BRFSS}. To quantify distributional similarity, we use the Jensen–Shannon (JS) divergence, which measures the difference between two probability distributions. Formally, for two distributions $P$ and $Q$, the JS divergence is defined as $JS(P \parallel Q) = \frac{1}{2} KL(P \parallel M) + \frac{1}{2} KL(Q \parallel M)$, where $M = \frac{1}{2}(P + Q)$ and $KL(P \parallel Q) = \sum_i P(i)\log_2\frac{P(i)}{Q(i)}$ denotes the Kullback--Leibler (KL) divergence, which quantifies how one probability distribution differs from another. A JS divergence value of 0 indicates identical distributions, while values approaching 1 indicate increasing dissimilarity.

We then evaluate the synthetic populations generated from these data. Again, we use JS divergence to assess agreement with the BRFSS ground truth distributions. In addition, we compute Pearson’s correlation coefficient ($r$) and examine spatial residual patterns relative to external benchmark datasets, using \textit{health insurance coverage} and \textit{general health status} as case studies. Synthesized health insurance rates are compared with estimates from the 2019–2023 five-year ACS \citep{acs2023}, while synthesized general health is compared with the 2025 CDC PLACES release \citep{cdcplaces}, which provides model-based estimates derived from 2023 BRFSS data and has been internally and externally validated \citep{zhang2015validation}. Together, these analyses provide both distributional and spatial validation of the synthetic populations.

\section{Results}\label{SEC:Results}

\subsection{LLM generated survey data}
\begin{figure}[htbp]
    \centering
    \includegraphics[width=1\textwidth]{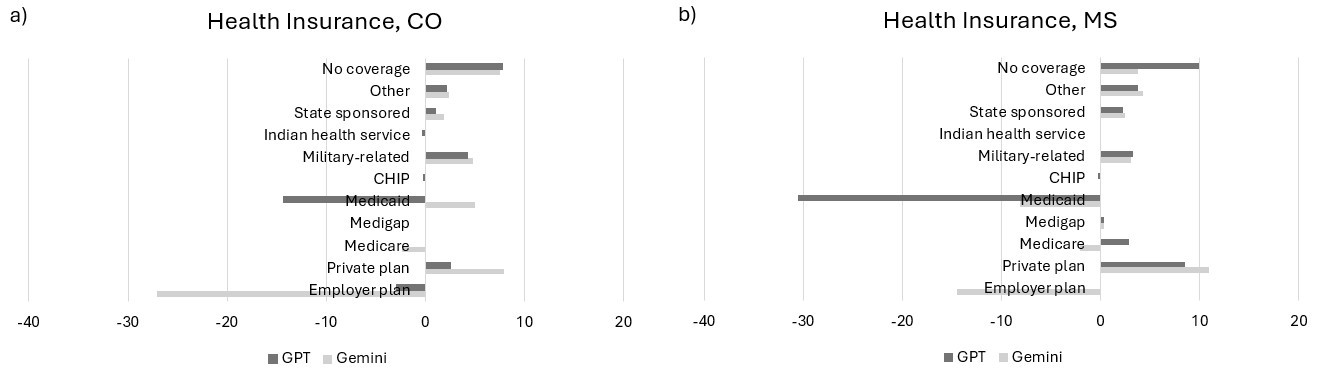}
    \caption{Residuals between ground truth and generated survey data for health insurance and general health. A negative residual indicates that the LLM overestimates the category relative to the ground truth data and a positive residual indicates that the LLM underestimates.}
    \label{fig:datagendiff}
\end{figure}

Table~\ref{tab:js_syn_shaded} shows the JS divergence between the LLM-generated and ground truth data, where a value of 0 indicates identical distributions. CO GPT has the lowest average divergence of $0.022$, followed by CO Gemini~($0.026$), MS Gemini~($0.031$), and MS GPT~($0.046$). Overall, MS distributions were harder for the models to reproduce than the CO distributions. Across the variables, \textit{sex} is reproduced almost perfectly with a mean of $=0.000$, followed by \textit{exercise} ($0.009$) and \textit{age} ($0.010$). The largest divergences are observed for \textit{smoker status} ($0.043$), \textit{income} ($0.045$), \textit{heart disease} ($0.049$), and \textit{insurance} ($0.129$). For example, with \textit{insurance} as shown in Figure~\ref{fig:datagendiff}, Gemini overestimates the share of the population with employer plans in both CO and MS, while GPT over-represents Medicaid plans, especially in MS. Both models significantly underestimate the share of the population with no health care coverage and private health insurance. 

\begin{table}[htbp]
\centering
\caption{JS divergence between ground truth and the LLM-generated survey data. Darker shading indicates larger divergence.}
\label{tab:js_syn_shaded}
\small
\begin{tabular}{lccccc}
\toprule
Variable & CO Gemini & CO GPT & MS Gemini & MS GPT & Row Mean \\
\midrule
Age            & \js{13}{0.010} & \js{11}{0.003} & \js{15}{0.018} & \js{13}{0.009} & \meanval{0.010} \\
BMI            & \js{15}{0.017} & \js{17}{0.024} & \js{13}{0.010} & \js{12}{0.008} & \meanval{0.015} \\
Depression     & \js{13}{0.010} & \js{13}{0.011} & \js{18}{0.028} & \js{19}{0.031} & \meanval{0.020} \\
Diabetes       & \js{23}{0.046} & \js{27}{0.061} & \js{26}{0.057} & \js{25}{0.053} & \meanval{0.054} \\
Education      & \js{15}{0.019} & \js{13}{0.010} & \js{13}{0.011} & \js{19}{0.033} & \meanval{0.018} \\
Exercise       & \js{10}{0.000} & \js{15}{0.017} & \js{10}{0.002} & \js{14}{0.016} & \meanval{0.009} \\
Flu Vaccination  & \js{14}{0.014} & \js{10}{0.001} & \js{10}{0.000} & \js{11}{0.005} & \meanval{0.005} \\
General Health & \js{17}{0.024} & \js{13}{0.012} & \js{12}{0.007} & \js{16}{0.021} & \meanval{0.016} \\
Heart Disease  & \js{23}{0.045} & \js{14}{0.014} & \js{32}{0.078} & \js{27}{0.060} & \meanval{0.049} \\
Income         & \js{17}{0.025} & \js{14}{0.014} & \js{20}{0.037} & \js{39}{0.102} & \meanval{0.045} \\
Insurance      & \js{47}{0.130} & \js{34}{0.086} & \js{44}{0.121} & \js{60}{0.178} & \meanval{0.129} \\
Race           & \js{16}{0.021} & \js{13}{0.011} & \js{20}{0.034} & \js{21}{0.039} & \meanval{0.026} \\
Sex            & \js{10}{0.000} & \js{10}{0.001} & \js{10}{0.000} & \js{10}{0.000} & \meanval{0.000} \\
Smoker         & \js{12}{0.006} & \js{23}{0.047} & \js{18}{0.027} & \js{36}{0.093} & \meanval{0.043} \\
\midrule
Column Mean    & \meanval{0.026} & \meanval{0.022} & \meanval{0.031} & \meanval{0.046} & \meanval{0.031} \\
\bottomrule
\end{tabular}
\end{table}

\begin{figure}[htbp]
    \centering
    \includegraphics[width=0.85\textwidth]{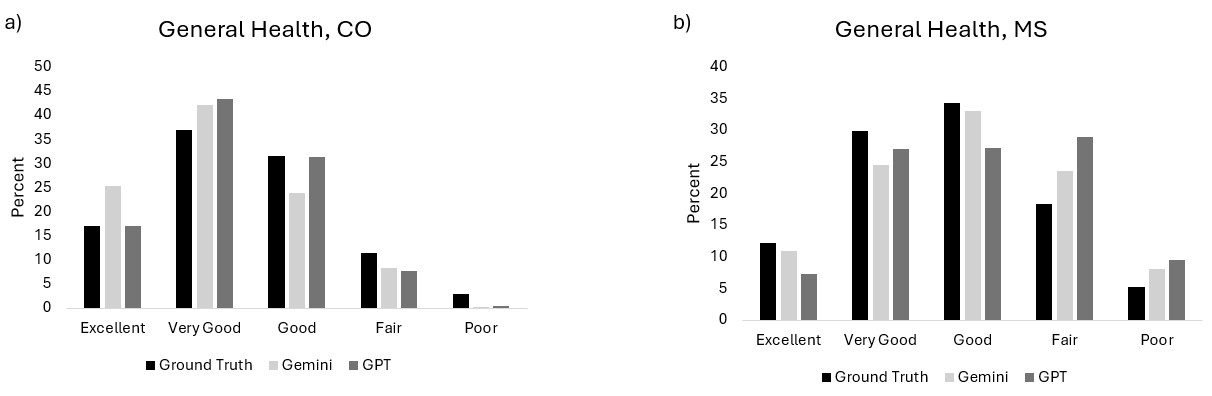}
    \caption{Distributions for CO and MS comparing ground truth and the LLM-generated individual-level data for general health.}
    \label{fig:genhealth}
\end{figure}

The ground truth data show that CO populations are more educated and physically active, have higher incomes, report better general health, and have lower BMI than MS populations (see descriptive statistics of the generated data in the Supplementary Materials, see \url{https://doi.org/10.17605/OSF.IO/H4STJ}. At the data-generation stage (Section \ref{SEC:Gen}), the LLMs generally capture these broad differences. For example, Figure~\ref{fig:genhealth} shows that both models reproduce the healthier profile in CO, although slightly exaggerated, and the less healthy profile in MS. 

\subsection{Population Synthesis}

\subsubsection{Variable Distributions}

After population synthesis, the differences in performance between the LLM-based populations become smaller as shown in Table~\ref{tab:js_synpop_shaded}. The average divergences for the LLM-based synthetic populations are $0.026$ for CO Gemini, $0.020$ for CO GPT, $0.021$ for MS Gemini, and $0.021$ for MS GPT. Relative to the survey data generated stage, this indicates that the IPF step reduces model-to-model differences overall, although not consistently for every variable. Overall, \textit{insurance} has the maintains the lowest mean divergence of $0.070$, followed by \textit{education} and \textit{income} (both $0.022$), \textit{BMI} ($0.021$), and \textit{flu vaccination} ($0.020$).

The BRFSS-based synthetic populations often captured the closest match to the ground truth, with column means of $0.007$ for both CO BRFSS and MS BRFSS (Table~\ref{tab:js_synpop_shaded}); they outperform the LLM-based populations for \textit{BMI}, \textit{diabetes}, \textit{heart disease}, \textit{insurance}, and \textit{race}. However, these should not be interpreted as a fully independent validation benchmark, since the BRFSS-based synthetic populations are generated using BRFSS-based inputs and then compared against BRFSS-derived target distributions. Any differences between the BRFSS-based synthetic populations and the target distributions are a function of the fitting and expansion process. 

\begin{table}[htbp]
\centering
\caption{JS divergence between the ground truth data and the synthetic populations. Darker shading indicates larger divergence.}
\label{tab:js_synpop_shaded}
\small
\begin{tabular}{lccccccc}
\toprule
Variable & CO BRFSS & CO Gemini & CO GPT & MS BRFSS & MS Gemini & MS GPT & Row Mean \\
\midrule
Age            & \js{10}{0.001} & \js{11}{0.002} & \js{10}{0.001} & \js{10}{0.001} & \js{10}{0.000} & \js{10}{0.001} & \meanval{0.001} \\
BMI            & \js{10}{0.001} & \js{17}{0.017} & \js{28}{0.043} & \js{10}{0.000} & \js{16}{0.015} & \js{30}{0.048} & \meanval{0.021} \\
Depression     & \js{10}{0.000} & \js{22}{0.030} & \js{12}{0.004} & \js{10}{0.000} & \js{13}{0.007} & \js{10}{0.001} & \meanval{0.007} \\
Diabetes       & \js{10}{0.000} & \js{27}{0.040} & \js{27}{0.040} & \js{10}{0.001} & \js{19}{0.021} & \js{13}{0.008} & \meanval{0.018} \\
Education      & \js{26}{0.038} & \js{20}{0.024} & \js{20}{0.023} & \js{24}{0.034} & \js{12}{0.005} & \js{12}{0.005} & \meanval{0.022} \\
Exercise       & \js{10}{0.000} & \js{10}{0.001} & \js{12}{0.004} & \js{10}{0.000} & \js{12}{0.004} & \js{12}{0.005} & \meanval{0.002} \\
Flu Vaccination & \js{27}{0.041} & \js{23}{0.032} & \js{10}{0.001} & \js{27}{0.040} & \js{10}{0.001} & \js{12}{0.006} & \meanval{0.020} \\
General Health & \js{10}{0.000} & \js{19}{0.021} & \js{22}{0.029} & \js{10}{0.000} & \js{11}{0.002} & \js{17}{0.017} & \meanval{0.012} \\
Heart Disease  & \js{10}{0.000} & \js{27}{0.040} & \js{16}{0.014} & \js{10}{0.000} & \js{30}{0.049} & \js{15}{0.011} & \meanval{0.019} \\
Income         & \js{15}{0.012} & \js{22}{0.030} & \js{16}{0.015} & \js{15}{0.012} & \js{24}{0.033} & \js{23}{0.032} & \meanval{0.022} \\
Insurance      & \js{10}{0.001} & \js{54}{0.107} & \js{40}{0.073} & \js{11}{0.002} & \js{57}{0.114} & \js{60}{0.121} & \meanval{0.070} \\
Race           & \js{10}{0.001} & \js{19}{0.021} & \js{12}{0.006} & \js{10}{0.001} & \js{21}{0.027} & \js{15}{0.011} & \meanval{0.011} \\
Sex            & \js{10}{0.000} & \js{10}{0.000} & \js{10}{0.000} & \js{10}{0.000} & \js{10}{0.000} & \js{10}{0.000} & \meanval{0.000} \\
Smoker         & \js{10}{0.001} & \js{12}{0.005} & \js{23}{0.031} & \js{10}{0.000} & \js{15}{0.011} & \js{20}{0.023} & \meanval{0.012} \\
\midrule
Column Mean    & \meanval{0.007} & \meanval{0.026} & \meanval{0.020} & \meanval{0.007} & \meanval{0.021} & \meanval{0.021} & \meanval{0.017} \\
\bottomrule
\end{tabular}
\end{table}

For several variables, the LLM-based populations perform similarly to the BRFSS-based populations with low divergence scores across the board (e.g. \textit{age}, \textit{exercise}, and \textit{sex}).
For \textit{income}, divergence is high across the board, pointing to limitations of IPF rather than the LLM-generated data. In a small number of cases, the LLM-based populations show lower JS divergence than the BRFSS-based ones. For example, MS Gemini and MS GPT both achieve lower JS divergence than MS BRFSS for \textit{education} ($0.005$ and $0.005$ versus $0.034$), and both also outperform MS BRFSS for \textit{flu vaccine} ($0.001$ and $0.006$ versus $0.040$).

The downstream effects of using these LLM-generated survey datasets for population-synthesis are mixed. Table~\ref{tab:js_diff_scores} presents the change in JS divergence for each variable after the IPF step.
Negative values indicate reduced divergence and improved fit to the ground truth, whereas positive values indicate increased divergence and worsened fit. In several cases, the synthetic populations are closer to the ground truth than the generated data. The largest improvements occur for MS GPT, including \textit{income} ($-0.070$), \textit{smoker status} ($-0.070$), \textit{insurance} ($-0.057$), \textit{heart disease} ($-0.049$), and \textit{diabetes} ($-0.045$). These variables were also among the weakest-performing variables in the data-generation stage, indicating that substantial upstream error does not necessarily propagate through IPF.

\begin{table}[htbp]
\centering
\caption{Difference in JS divergence scores after the IPF step. Negative values indicate better fit while darker shading indicates larger magnitude of change.}
\label{tab:js_diff_scores}
\small
\begin{tabular}{lcccc}
\toprule
Variable & CO Gemini & CO GPT & MS Gemini & MS GPT \\
\midrule
Age            & \improve{22}{-0.008} & \improve{10}{-0.002} & \improve{38}{-0.018} & \improve{22}{-0.008} \\
BMI            & 0.000                & \worsen{40}{0.019}  & \worsen{16}{0.005}   & \worsen{70}{0.040} \\
Depression     & \worsen{42}{0.020}   & \improve{20}{-0.007} & \improve{44}{-0.021} & \improve{60}{-0.030} \\
Diabetes       & \improve{18}{-0.006} & \improve{44}{-0.021} & \improve{66}{-0.036} & \improve{75}{-0.045} \\
Education      & \worsen{16}{0.005}   & \worsen{30}{0.013}  & \improve{18}{-0.006} & \improve{56}{-0.028} \\
Exercise       & \worsen{8}{0.001}    & \improve{30}{-0.013} & \worsen{10}{0.002}   & \improve{26}{-0.011} \\
Flu Vaccination  & \worsen{38}{0.018}   & 0.000               & \worsen{8}{0.001}    & \worsen{8}{0.001} \\
General Health & \improve{12}{-0.003} & \worsen{36}{0.017}  & \improve{16}{-0.005} & \improve{14}{-0.004} \\
Heart Disease  & \improve{16}{-0.005} & 0.000               & \improve{58}{-0.029} & \improve{80}{-0.049} \\
Income         & \worsen{16}{0.005}   & \worsen{8}{0.001}   & \improve{14}{-0.004} & \improve{90}{-0.070} \\
Insurance      & \improve{48}{-0.023} & \improve{30}{-0.013} & \improve{20}{-0.007} & \improve{85}{-0.057} \\
Race           & 0.000                & \improve{16}{-0.005} & \improve{20}{-0.007} & \improve{56}{-0.028} \\
Sex            & 0.000                & \improve{8}{-0.001}  & 0.000                & 0.000 \\
Smoker         & \improve{8}{-0.001}  & \improve{34}{-0.016} & \improve{34}{-0.016} & \improve{90}{-0.070} \\
\bottomrule
\end{tabular}
\end{table}

\begin{figure}[b]
    \centering
    \includegraphics[width=1\textwidth]{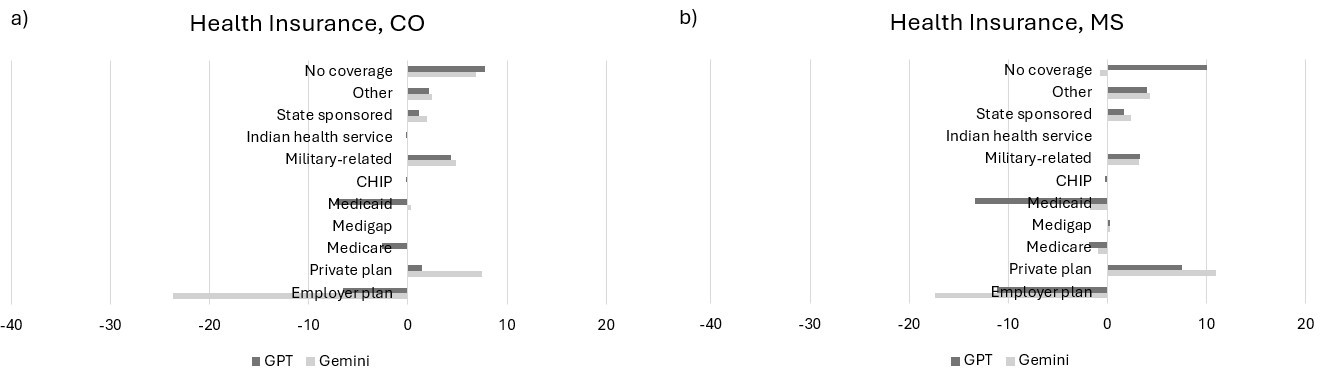}
    \caption{Residuals between the ground truth and the synthetic populations for \textit{health insurance} and \textit{general health}. A negative or positive residual indicates that the LLM overestimates or underestimates respectively. }
    \label{fig:pop_synth}
\end{figure}

\textit{Insurance} provides a clear example of the mixed effects of population synthesis. Overall, the synthetic population is closer to the ground truth. This is likely because the overestimation of the \textit{employer-plan} and \textit{Medicaid} categories reduced in the synthetic population, especially for GPT (Figure \ref{fig:pop_synth} a and b). By contrast, the error for the \textit{uninsured} category persists for GPT because the generated data contained no uninsured individuals and thus the synthesis process cannot recover that part of the distribution. 

IPF can worsen variables that were captured reasonably well at the data-generation stage. For example, \textit{BMI} worsens for MS GPT ($+0.040$) and CO GPT ($+0.019$). Importantly, although in the data-generation stage, the state differences in BMI between CO and MS are clear, they become slightly less pronounced after population synthesis. Finally, the fitted variables do not uniformly show improved agreement after IPF, although several do. For example, \textit{income} and \textit{education} worsen for CO, despite being included in the fitting process. This suggests that even fitted marginals may shift in unexpected ways after expansion, particularly for correlated variables such as these.

\subsubsection{Spatial Distributions}

\begin{figure}[htbp]
    \centering
    \includegraphics[width=0.9\textwidth]{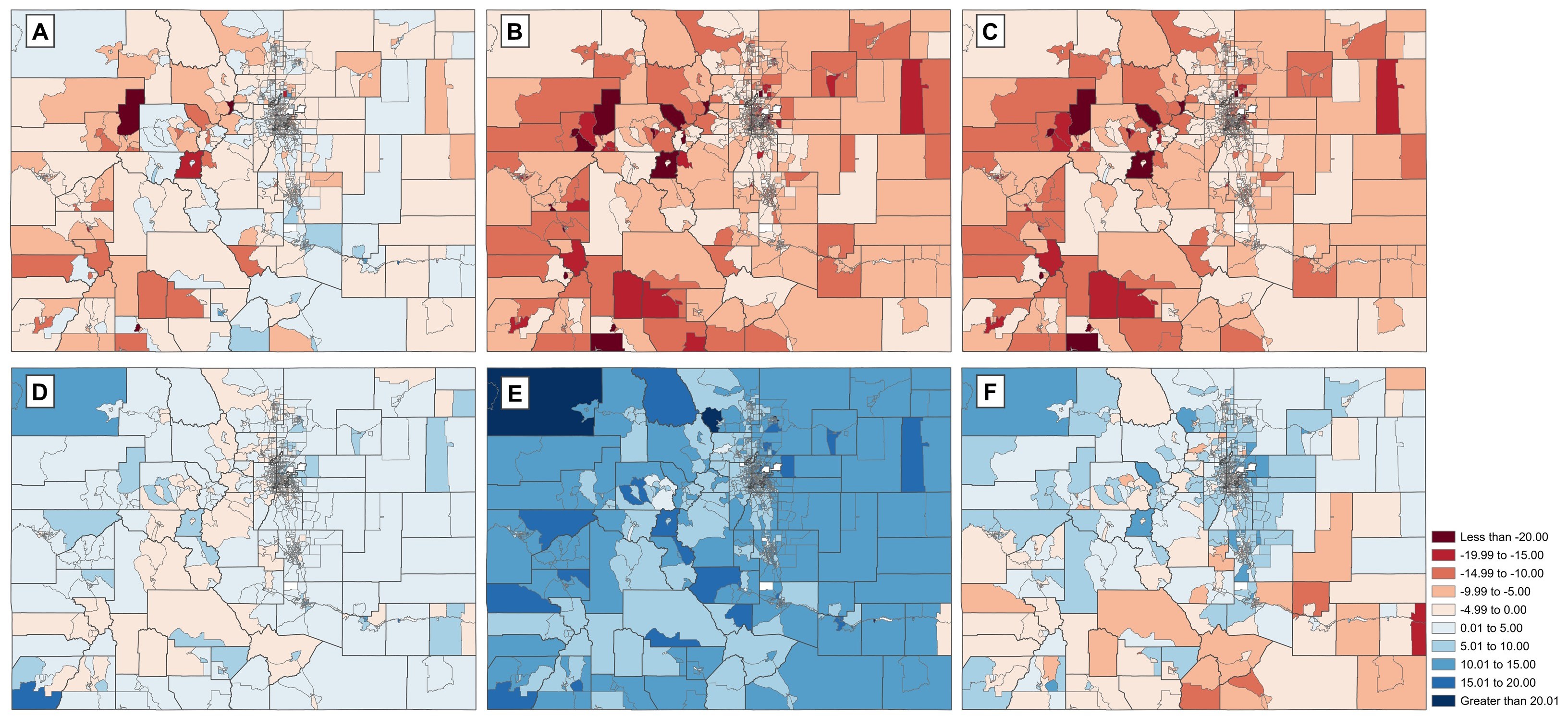}
    \caption{Spatial distribution of residuals for \textit{health insurance} and \textit{general health} across CO census tracts. Maps (A–C) show residuals for health insurance coverage compared against ACS 2023 estimates for: (A) BRFSS-based; (B) GPT-based; and (C) Gemini-based synthetic populations. Maps (D–F) show residuals for poor health status compared against 2023 CDC PLACES estimates for: (D) BRFSS-based; (E) GPT-based; and (F) Gemini-based synthetic populations. }
    \label{fig:co_residuals}
\end{figure}

\begin{figure}[htbp]
    \centering
    \includegraphics[width=0.75\textwidth]{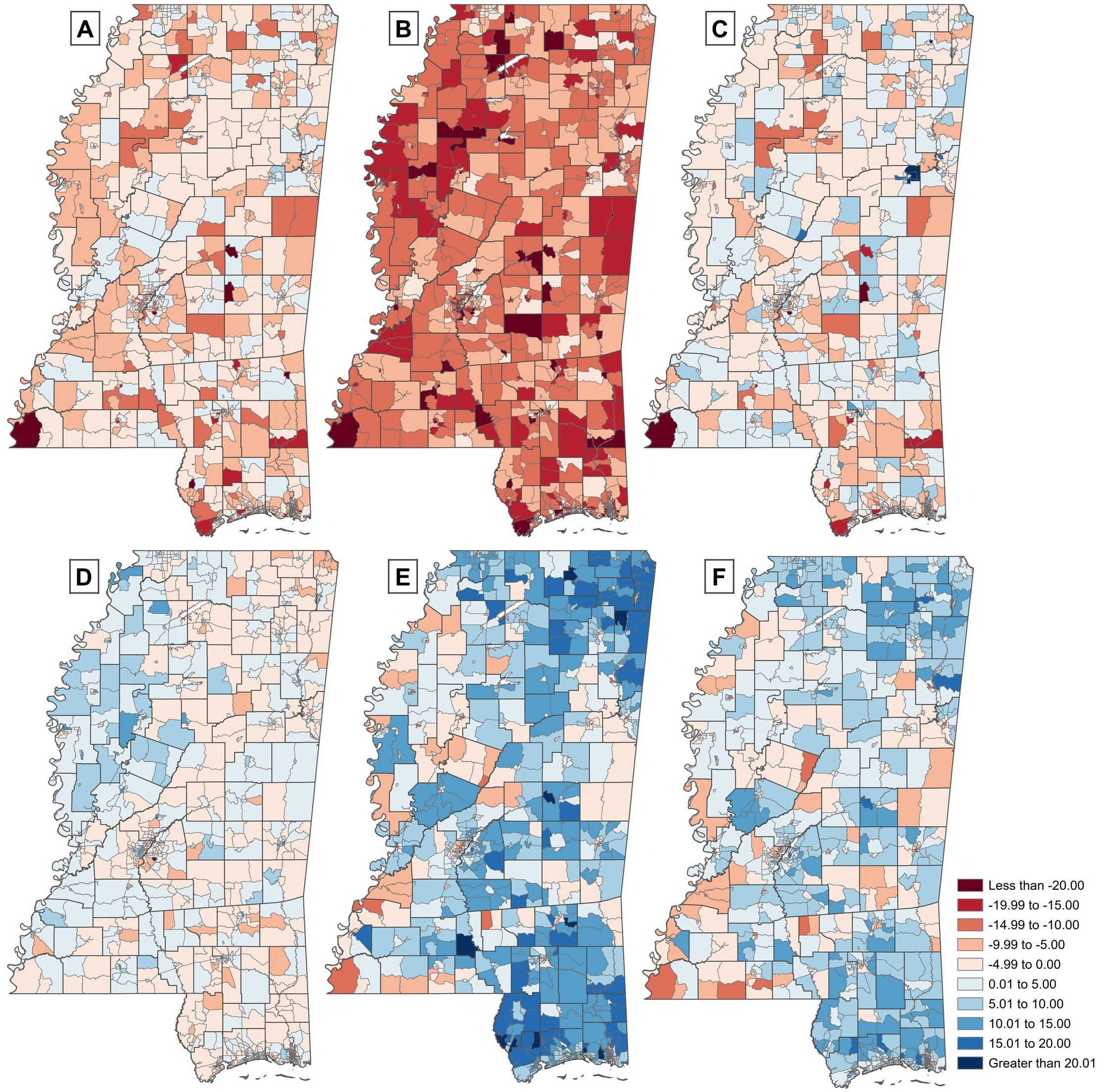}
    \caption{Spatial distribution of residuals for health insurance and general health across Mississippi census tracts. Maps (A–C) show residuals for health insurance coverage compared against ACS 2023 estimates for: (A) BRFSS-based; (B) GPT-based; and (C) Gemini-based synthetic populations. Maps (D–F) show residuals for poor health status compared against 2023 CDC PLACES estimates for: (D) BRFSS-based; (E) GPT-based; and (F) Gemini-based synthetic populations.}
    \label{fig:ms_residuals}
\end{figure}

To assess how well the synthetic populations reproduce geographic variation, we aggregated each population to the census tract level and computed the percent insured and percent fair or poor general health. These small-area estimates (SAEs) were compared to external benchmarks: ACS estimates for insurance coverage and CDC PLACES estimates for general health. In addition to residual maps (Figures~\ref{fig:co_residuals} and \ref{fig:ms_residuals}), we quantify spatial agreement using Pearson’s $r$ between synthetic and ground truth estimates.

For CO \textit{insurance coverage} (Figure \ref{fig:co_residuals} A--C), the BRFSS-based synthetic population shows the closest agreement with ACS benchmarks, with the smallest residuals and moderate spatial correlation ($r=0.592$). In contrast, both LLM-based populations systematically overestimate coverage, producing predominantly negative residuals and weak spatial agreement (Gemini: $r=0.216$; GPT: $r=-0.084$). This reflects earlier distributional errors in the data-generation stage and illustrates how they propagate into the synthetic populations. For CO \textit{general health} (Figure \ref{fig:co_residuals} D--E), the BRFSS-based population again performs best, with small residuals and strong agreement with CDC PLACES estimates ($r=0.830$), capturing geographic heterogeneity well. The LLM-based populations tend to underestimate fair or poor health, consistent with their bias toward better health, resulting in more positive residuals—especially for GPT. Despite this, spatial correspondence remains relatively strong (GPT: $r=0.773$; Gemini: $r=0.744$), indicating that the LLM-based synthetic populations still capture the counties with relatively better or worse health, even when the overall level is biased towards better health.

MS shows similar but slightly stronger patterns overall. For fair or poor \textit{general health}, both LLM-based synthetic populations align well with the CDC PLACES estimates, and in this case Gemini performs particularly well (Figure \ref{fig:ms_residuals} D--F). Although the residuals for Gemini are comparable to the BRFSS-based synthetic population, the spatial correlation is the highest of the three approaches ($r=0.874$ for Gemini, compared with $r=0.832$ for BRFSS and $r=0.857$ for GPT). GPT also performs strongly, although its maps show more areas where fair or poor health is underestimated. These results suggest that, in MS, the LLM-based synthetic populations preserve the relative spatial patterning of poor health quite well. 

For MS \textit{insurance coverage}, however, the BRFSS-based synthetic population remains more accurate than the LLM-based alternatives (Figure \ref{fig:ms_residuals} A--C). It has lower residuals overall and a moderate correlation with ACS estimates ($r=0.476$). As in CO, both GPT- and Gemini-based synthetic populations overestimate the insured share, reflecting errors introduced earlier in the data-generation stage. As a result, their agreement with the ACS spatial pattern is weak ($r=0.025$ for GPT and $r=0.264$ for Gemini). Thus, while the LLM-based synthetic populations show some ability to preserve spatial heterogeneity for \textit{general health}, they perform poorly for \textit{insurance coverage} in both states.

Overall, BRFSS-based synthetic populations provide the most accurate tract-level estimates across outcomes. Among LLM-based populations, performance is outcome-dependent: \textit{general health} is reproduced reasonably well spatially—especially in MS—while \textit{insurance coverage} remains poorly captured due to biases introduced during data generation.

\section{Discussion and Conclusion}\label{SEC:Discussion}
This study has examined the potential of zero-shot LLM-generated BRFSS-like survey data as an input to a conventional IPF workflow for geographically explicit population synthesis. Overall, the results show promise, but not a complete answer. Even without being given target proportions, both GPT-4.1 and Gemini-2.5-Pro reproduced several broad state-level differences between CO and MS, and many of these differences carried through to the tract-level synthetic populations. The results also show clear limits to what zero-shot generation can achieve. The accuracy of the generated survey data is strongly variable-dependent, and good agreement at the survey-data stage does not guarantee accurate downstream synthetic populations.

One important finding from our work is that the LLMs generally captured the direction of the major cross-state contrasts. This is encouraging as it suggests that the models are not simply generating generic U.S. adults that we might observe in national data, but can produce state-conditioned records that reflect meaningful geographic context. However, this success was uneven. Variables such as \textit{sex}, \textit{age}, and \textit{exercise} were reproduced with low divergence, whereas \textit{insurance}, \textit{heart disease}, \textit{diabetes}, \textit{income}, and \textit{smoker status} were consistently harder to match. These more difficult variables tend to have a large number of and/or less observed categories.

A second key finding of our work is that model performance at the survey-data stage does not map cleanly onto performance after population synthesis. CO GPT had the lowest average divergence among the generated datasets, while MS GPT had the highest. After the IPF step, however, the two LLMs looked much more similar overall. This convergence indicates that the fitting and expansion process can partially regularize differences between the generated survey datasets. At the same time, IPF did not simply ``fix'' the generated data. For some variables, divergence decreased substantially after synthesis, especially for the weaker MS GPT inputs; for others, divergence increased. This mixed downstream behavior is substantively important. It shows that evaluating LLM-generated survey data only would provide an incomplete picture of their practical utility in population synthesis. 

Surprisingly, there were cases in which an LLM-based synthetic population outperforms a BRFSS-based one. The improved performance of some LLM-based populations for \textit{education} and \textit{flu vaccination} may reflect smoothing of noisy survey distributions in real data, better alignment with the fitted controls, or artifacts of the expansion procedure. Regardless of the mechanism, these cases highlight that the relationship between input realism and downstream fit is more complicated than a simple ``real data always wins'' narrative. They reinforce the need to evaluate generated survey data in the context of the exact synthesis procedure and validation target for which they will be used. However, our findings do suggest a practical role for zero-shot LLM-generated survey data, but not yet as a drop-in replacement for high-quality survey data. The strongest use case appears to be as a supplementary or provisional data source. In contrast, fully unconstrained zero-shot generation appears too unreliable for variables that are policy-sensitive, rare, or structurally complex. For these variables, stronger guidance during generation or stronger post-generation correction is likely necessary.

As with all research, our research has several limitations which should be noted. First, the study considers only two states, two LLMs, and one population-synthesis framework. The extent to which these findings generalize to other geographies, models, or synthesis approaches is unknown. Second, the prompting strategy is intentionally minimal and zero-shot; this is useful for testing baseline capability, but it likely underestimates what could be achieved with stronger guidance during generation. Third, our evaluation focuses primarily on categorical marginals and two downstream case studies. Although these are appropriate for a first assessment, future work should examine higher-order joint distributions, uncertainty, and sensitivity of downstream simulation outcomes. Fourth, the BRFSS is a prominent public survey, so some of its structure is likely reflected in LLM  pretraining data; future validation work should therefore place particular emphasis on surveys, locations, or time periods that are less likely to have been seen by the models. Even with these limitations and areas of further work, our results suggest that zero-shot LLM-generated survey data can provide a useful starting point for spatial population synthesis in the US public health domain, especially when conventional survey data are limited or unavailable. However, their accuracy depends on the variable and the geography and changes from survey generation to population synthesis, making careful validation of the synthetic populations themselves essential before they are used in substantive modeling or policy analysis.

\section*{ACKNOWLEDGMENTS}
This research was funded by NSF Award No. 2109647 and 2302970.

\footnotesize
\bibliographystyle{plainnat}
\bibliography{demobib}

\section*{AUTHOR BIOGRAPHIES}
\noindent {\bf \MakeUppercase{Taylor Anderson}} is an Associate Professor in the Department of Geography and Geoinformation Science (GGS) at George Mason University (GMU). Her research focuses on modeling the spread of diseases in human and ecological systems. Her e-mail address is \email{tander6@gmu.edu} and her website is \url{https://science.gmu.edu/directory/taylor-anderson}.\\

\noindent {\bf \MakeUppercase{Sara Von Hoene}} is an undergraduate researcher pursuing a B.S. in Geography with a concentration in Geoinformatics in the Department of GGS at GMU. Her research focuses on geospatial analysis. Her email address is \email{svonhoen@gmu.edu}.\\

\noindent {\bf \MakeUppercase{Orhan Yagizer Cinar}} is an undergraduate pursuing a B.S. in the Department of Computer Science at GMU and a research assistant in GGS. His research interests include LLMs and AI-driven healthcare solutions. His email address is \email{ocinar@gmu.edu} and his website is \url{https://orhancinar.com}.\\

\noindent {\bf \MakeUppercase{Emma Von Hoene}} is a Ph.D. candidate in the Department of GGS at GMU. Her research focuses on data-driven modeling and geospatial analysis. Her email address is \email{evonhoen@gmu.edu}.\\

\noindent {\bf \MakeUppercase{Amira Roess}} is a Professor of Epidemiology and Global Health at GMU's College of Public Health. Her research uses epidemiology and evaluation methods to study human, wildlife, pet and environmental determinants of health. Her email address is \email{aroess@gmu.edu} and her website is \url{https://publichealth.gmu.edu/profiles/aroess}.\\

\noindent {\bf \MakeUppercase{Andrew Crooks}} is a Professor in the Department of Geography at the University of Buffalo. His research interests include GIS and agent-based modeling. His email address is \email{atcrooks@buffalo.edu} and his website is \url{https://www.gisagents.org/}.\\

\noindent {\bf \MakeUppercase{Hamdi Kavak}} is an Associate Professor in the Department of Computational and Data Sciences and Co-Director of Center for Social Complexity at GMU. His research combines data science with modeling and simulation to investigate challenges in urban and social systems. His email address is \email{hkavak@gmu.edu} and his website is \url{https://hamdikavak.com/}.\\

\end{document}